\documentclass[]{JHEP3}

\title{\bf Holographic currents in first order Gravity and finite
Fefferman-Graham expansions}

\author{M\'aximo Ba\~nados$^a$, Olivera Mi\v{s}kovi\'c$^{a,b}$
and Stefan Theisen$^c$\\
$^a$Departamento de F\'\i sica, P. Universidad
Cat\'olica de Chile,\\ Casilla 306, Santiago 22, Chile\\
$^b$Instituto de F\'{i}sica, P. Universidad Cat\'{o}lica
de Valpara\'{\i}so,\\ Casilla 4059, Valpara\'{\i}so, Chile\\
$^c$Max-Planck-Institut f\"ur
gravitationsphysik, Albert-Einstein-Institut,\\ 14476 Golm, Germany \\
E-mail: \email{maxbanados@fis.puc.cl},
\email{olivera.miskovic@ucv.cl},\\ \email{theisen@aei.mpg.de}}

\abstract{ We study the holographic currents associated to
Chern-Simons theories. We start with an example in three
dimensions and find the holographic representations of vector and
chiral currents reproducing the correct expression for the chiral
anomaly. In five dimensions, Chern-Simons theory for AdS group
describes first order gravity and we show that there exists a
gauge fixing leading to a finite Fefferman-Graham expansion. We
derive the corresponding holographic currents, namely, the stress
tensor and spin current which couple to the metric and torsional
degrees of freedom at the boundary, respectively. We obtain the
correct Ward identities for these currents by looking at the bulk
constraint equations.}

\begin{document}

\section{Introduction}

The AdS/CFT correspondence
\cite{Maldacena,Gubser-Klebanov-Polyakov,Witten98} has uncovered a
deep and still somewhat mysterious relationship between fields
propagating in ($d+1$)-dimensional anti-de Sitter (AdS) space and
correlators in a $d$-dimensional Conformal Field Theory (CFT).
Scalars, vectors, spinors and tensor fields on AdS have been
studied (see\cite{Aharony-Gubser-Maldacena-Ooguri-Oz,Skenderis}
for reviews) and the expected results are obtained in all cases.

One of the most studied examples of this correspondence has been
the gravitational field.  This bulk field is dual to the CFT's
energy momentum tensor, and solving the bulk Einstein equations in
the presence of a cosmological term allows the computation of
energy momentum tensor correlators. The calculation of
the holographic anomalies performed in \cite{Henningson-Skenderis}
has provided strong support for the validity of the
correspondence.

In this paper we shall analyze the holographic structure of two
unrelated, although similar, systems. In section 2, we study
Abelian and non-Abelian Chern-Simons theories in three dimensions
and work out the 1-point functions and anomalies associated to
gauge symmetries of a dual theory on the boundary. The aim of section 2
is to put a simple Chern-Simons system in the AdS/CFT language and use this
example as a warm-up exercise to deal with the more complicated
case of five-dimensional first order Chern-Simons gravity,
discussed in Sec. 3. Five dimensional Chern-Simons gravity has more degrees
of freedom than standard gravity due to the fact that the spin connection is
dynamical \cite{Banados-Garay-Henneaux}.  We carry out the holographic AdS/CFT
prescription associated to this system and extract the vevs and corresponding
Ward identities associated to the metric and spin connection degrees of freedom.

The systems discussed in sections 2 and 3 both have finite
Fefferman-Graham (FG) \cite{Fefferman-Graham} expansions.  The
first example of a finite FG expansions appeared in
\cite{Skenderis-Solodukhin} in the context of three-dimensional
gravity (see also \cite{Banados-Chandia-Ritz,Banados-Caro} for a
Chern-Simons formulation). Incidentally, in three dimensions, the
connection between anti-de Sitter space and the full 2d conformal
group has been known for a long time \cite{Brown-Henneaux}.  This
correspondence was later reformulated in terms of the Chern-Simons
formulation \cite{Achucarro-Townsend,Witten88} of gravity, and
associated WZW theories
\cite{Witten89,Coussaert-Henneaux-vanDriel,Banados-Bautier-Coussaert-Henneaux-Ortiz}.
For a recent complete review of these issues and their
applications, see \cite{Carlip}.

\section{Chiral anomaly in two dimensions \label{U1xU1}}

Before going into the more complicated case of first order gravity
in five dimensions, let us review here the holographic description
of the chiral anomaly in two dimensions, which is perhaps  the
simplest application of the AdS/CFT ideas. It also provides a
simple example in which the FG expansion is finite. Most of the
material of this section is well-known in different contexts.

In even dimensions, gauge fields can be coupled to a vector
current or axial current.  It is well known that only one of these
gauge symmetries can be preserved at the quantum level. In the
particular case of two dimensions, the vector $(J^i)$ and axial
$(J_5^i)$ currents are related by
\begin{equation}\label{cf}
J_5^i = \varepsilon^{ij} J_j\,.
\end{equation}
In the presence of a non-zero $U(1)$ gauge field $A_i$, it follows
that
\begin{equation}\label{id0}
\langle \partial _i J^i  \rangle=0\,, \qquad \langle
\partial _i J^i_5 \rangle = F\,,
\end{equation}
where $F$ is the field strength associated to $A$.

Chern-Simons theory in three dimensions provides a natural arena
to describe (\ref{id0}) holographically. Since the AdS/CFT dual of
a conserved current is a gauge field, one may naively think that a
single Abelian Chern-Simons theory may be the right description
leading to (\ref{id0}). This is not the case, and one needs two
Abelian fields. We first explain why a single Chern-Simons theory
is not enough to represent the chiral anomaly.

Consider an Abelian Chern-Simons action $I[B]=\kappa\int_{M_3} B
dB$. (In the whole text we omit the wedge product symbol.) This action is
not gauge invariant for parameters with non-zero support at the
boundary $\partial M_3=M_2$. In fact, under $\delta B = d\lambda$, the action changes
as $\delta I = 2\kappa\int_{M_2} \lambda F $ with $F=dB$.
Naively, one may conclude that this simple
calculation represents the holographic version of the chiral
anomaly, in analogy with the five-dimensional case discussed in
\cite{Witten98}. This is however not correct because there is a
mismatch between the number of sources.

According to AdS/CFT, the bulk fields boundary data (independent
fields at the boundary) plays the role of sources in the CFT.  For
a single Chern-Simons theory, the boundary data is parameterized
by only one component of the gauge field. In fact, let us choose
local coordinates on $M_{3}$ as $x^{\mu }=\left( \rho
,x^{i}\right) $, where the light-cone coordinates
$x^{i}=\left(x^{+},x^{-}\right) $ parameterize the boundary
$M_{2}$ placed at $\rho =0$. The gauge field has components $B_\mu
= (B_\rho,B_+,B_-)$. In the gauge $B_\rho=0$, for example, the
fields $B_+,B_-$ become $\rho$-independent. The remaining equation
of motion $F_{+-}=0$ allows to solve $B_-$ as a function of $B_+$.
The field $B_+$ is thus the only ``source". On the other hand, for
a fermion coupled to a gauge field, the source is a vector $A_i$
with two components, thus it is clear that we need to double the
number of degrees of freedom.\footnote{In fact, the U(1)
Chern-Simons action is dual to a single chiral fermion  on the
boundary and it reproduces the chiral anomaly. However, in this
case the boundary theory does not have (preserved) symmetries and
we are not interested in these cases here.}

\subsection{Abelian chiral anomaly}

Consider then two Abelian Chern-Simons actions with fields $B$ and
$\bar B$,
\begin{equation}
I\left[ B,\bar{B}\right] = \kappa\int BdB - \kappa\int \bar B d\bar B
+  \Omega\left[ B,\bar{B}\right] \,,  \label{CS Abelian}
\end{equation}
where $\Omega$ is a boundary term to be fixed. The boundary data
of this action contains two fields, as desired.

The bulk local symmetries of this action are $U(1)\times U(1)$
gauge transformations acting independently on each field. Of
course,  these symmetries are broken at the boundary. The
interesting point is that the boundary term $\Omega$ can be chosen
such that {\it half} of the gauge symmetries of (\ref{CS Abelian})
are extended to the boundary. The unbroken part of the symmetries
will be related to the vector current, and the broken half
represents the axial current. Let us see how this is implemented
in the language of AdS/CFT.

The crucial point is the choice of the boundary term $\Omega$. We
choose
\begin{equation}
\Omega\left[ B,\bar{B}\right] =2\kappa
\int\limits_{M_{2}}d^{2}x\,\left(
B_{+}B_{-}+\bar{B}_{-}\bar{B}_{+}-2B_{+}\bar{B}_{-}\right).
\label{B}
\end{equation}
The action $I$ in (\ref{CS Abelian}), with this choice of boundary term, has the following properties.

First, consider the variation of (\ref{CS Abelian}) under gauge
transformations $ \delta B=d\lambda $ and $\delta
\bar{B}=d\bar{\lambda}$. One obtains
\begin{equation}\label{dI}
\delta I  =4\kappa \int\limits_{M_{2}}\left( \lambda -
\bar{\lambda}\right) (\partial _+ \bar B_- -\partial _- B_+  )\,.
\end{equation}
The action becomes strictly invariant --even at the boundary--
under the diagonal subgroup with $\bar{\lambda}=\lambda $.
Redefining $\lambda + \bar\lambda = \sigma_+$ and $\lambda -
\bar\lambda = \sigma_-$, we conclude that half of the gauge
symmetries, $\sigma_+$, of the action (\ref{CS Abelian}) are
preserved at the boundary, as desired.

Second, under generic variations of the fields, the action
(\ref{CS Abelian}) varies as
\begin{equation}\label{dIonshell}
\delta I=4\kappa \int\limits_{M_{2}}d^{2}x\,\left[ \delta
B_{+}\left( B_{-}- \bar{B}_{-}\right) +\delta \bar{B}_{-}\left(
\bar{B}_{+}-B_{+}\right) \right] + \mbox{e.o.m.}\,,
\end{equation}
where e.o.m. represents terms proportional to the equations of
motion $dB=0$ and $d\bar B=0$.  This shows that the bulk action
has an extremum, provided $B_+$ and $\bar B_-$ are fixed at the
boundary. The boundary fields $(B_+,\bar{B}_-)$ are thus the  two ``sources" in
this theory. We identify them with the 2d gauge field \footnote{A
similar phenomenon occurs in three-dimensional gravity and the
corresponding AdS/CFT interpretation. In the Chern-Simons
formulation, there are two $SO(2,1)$ gauge  fields $A$ and $\bar
A$. The induced 2-dimensional vielbein in the FG expansion is
constructed with one leg in $A$ and the other in $\bar A$. For
more details in this construction, see
\cite{Banados-Chandia-Ritz,Banados-Caro}.}
\begin{equation}\label{source} (A_+,A_-) = (B_+,\bar B_-)\,.
\end{equation}
Note now that the variation of the action under the broken
symmetry, eq. (\ref{dI}), becomes proportional to the gauge field curvature
since $\partial _+ \bar B_- -\partial _- B_+  = \partial _+ A_- -\partial _- A_+ =F$.
This is consistent with the chiral anomaly.

Third, we can now write the on-shell action as a function of the
sources, and compute vacuum expectation values of the currents. We
can compute directly the vector current expectation value
\begin{eqnarray}
\langle J^i   \rangle &=& {\delta I \over \delta A_i}\nonumber \\
    &=& \left({\delta I \over \delta B_+}, {\delta I \over \delta \bar B_- } \right)\\
    &=& 4\kappa \left( B_- - \bar B_-\, , \bar B_+ - B_+ \, \right)\,,\nonumber
\end{eqnarray}
where $I$ is the on-shell value of the action. Here, the first
line is the AdS/CFT definition of the expectation value for the
current. The second line follows from (\ref{source}), and the last
line from (\ref{dIonshell}). $B_-$ and $\bar B_+$ are not
independent of the sources $B_+$ and $\bar B_-$. From the
Chern-Simons bulk equations of motion, $dB=0$ and $d\bar B=0$, one
finds the non-local relations
\begin{equation}
B_{-}=\frac{\partial _{-}}{\partial _{+}}\,B_{+}\,,\qquad \bar{B
}_{+}=\frac{\partial _{+}}{\partial _{-}}\,\bar{B}_{-}\,.
\label{non-local solution}
\end{equation}
The final expression for the expectation value of the vector
current in the presence of the source $A_i$ is
\begin{equation}
\langle J^i   \rangle = 4\kappa \left( \frac{\partial
_{-}}{\partial _{+}} A_+ - A_-\, ,\,\frac{\partial _{+}}{\partial
_{-}} A_- - A_+ \right)\,.
\end{equation}
The corresponding formula for the axial current follows from (\ref{cf}),
\begin{equation}
\langle J^i_5   \rangle = 4\kappa \left( \frac{\partial
_{-}}{\partial _{+}} A_+ - A_-\, ,\, -\frac{\partial
_{+}}{\partial _{-}} A_- + A_+ \right)\,.
\end{equation}
We can finally check explicitly the two relations
\begin{eqnarray}
\partial _{i}\left\langle J^{i}\right\rangle &=&0\,,  \label{vector} \\
\partial _{i}\left\langle J_5^{i}\right\rangle &=&-8\kappa \left(
\partial _{+}A_{-}-\partial _{-}A_{+}\right)=-4\kappa \,\varepsilon
^{ij}F_{ij}\,,  \label{axial-vector}
\end{eqnarray}
in full consistency with the Dirac fermion expectation values.
Note that the Chern-Simons coupling then becomes related to the
electric charge as $q/2\pi =4\kappa $.

\subsection{Non-Abelian anomaly}

The above result can be generalized to a non-Abelian CS theory
invariant under $G\times G$, with $G$ semi-simple for simplicity.
Each of two sets of generators $G_{a}$ and $\bar{G}_{a}$ forms a
Lie algebra with structure constants ${f_{ab}}^c$. The two
isomorphic algebras commute with each other. The action with the
fundamental fields $B=B^{a}G_{a}$, $
\bar{B}=\bar{B}^{a}{\bar{G}}_{a}\,$ and the Cartan metric $
g_{ab}=\left\langle {G}_{a}{G}_{b}\right\rangle $ (that raises and
lowers group indices), is
\begin{equation}
I\left[ B,\bar{B}\right] =I_{CS}\left[ B\right] -I_{CS}\left[
\bar{B}\right] +2\kappa \int\limits_{M_{2}}d^{2}x\,\left(
B_{+}^{a}B_{-a}+\bar{B}_{+}^{a}
\bar{B}_{-a}-2B_{+}^{a}\bar{B}_{-a}\right) \,,  \label{CS
non-Abelian}
\end{equation}
where the CS action is
\begin{equation}
I_{CS}\left[ B\right] =\kappa \int\limits_{M_{3}}\left(
B^{a}F_{a}-\frac{2}{3}\,f_{abc}\,B^{a}B^{b}B^{c} \right) \,,
\label{3d CS action}
\end{equation}
and the boundary term is a covariant generalization of (\ref{B})
\cite{Arcioni-Blau-O'Loughlin}. The asymptotic conditions are
\begin{equation}
\left. B_+^a\right\vert _{\rho =0}=A_+^a\,,\qquad \left.
\bar{B}_-^a\right\vert _{\rho =0}=A_-^a\,, \label{boundary
condition}
\end{equation}
and $A^a=B_+^a\, dx^+ +\bar{B}_-^a \,dx^-$ is a fixed vector field
on $M_2$. The equations of motion
\begin{equation}
F^a\equiv dB^a+\frac{1}{2}\,{f_{bc}}^a\,B^b B^c = 0\,,\qquad \bar{
F}^a\equiv d\bar{B}^a + \frac{1}{2}\,{f_{bc}}^a\,\bar{B}^b\bar{B}
^c = 0\,, \label{F=0}
\end{equation}
give $B_-^a$ and $\bar{B}_+^a$ as non-local functions of $A^a$.
Similarly to the Abelian case, the action (\ref{CS non-Abelian})
is asymptotically invariant under the diagonal subgroup
$G_{D}\subset G\times G$ generated by ${G}_{a}+{\bar{G}}_{a}$ ,
while the other half of symmetries, generated by
${G}_{a}-{\bar{G}}_{a}$, are broken on the boundary. This
asymptotic structure of the gauge theory maps holographically to
the quantum structure of a field theory at the boundary, leading
to one preserved quantum current and another current possessing a
non-Abelian anomaly. Varying the action (\ref{CS non-Abelian})
on-shell, we find these currents as
\begin{equation}
\left\langle J_a^i\right\rangle =\frac{\delta I[A]}{ \delta
A_i^a}=2\kappa \,g_{ab}\,\varepsilon ^{ij}\left(
B_j^b-\bar{B}_j^b\right) \,,  \label{current 1}
\end{equation}
where $I$ is evaluated on-shell and $\bar{J}_a^i=\varepsilon ^{ij}
J_{ia}$. Using the field equations (\ref{F=0}), the covariant
derivatives of these currents are
\begin{eqnarray}
D^{(0)}_i\left\langle J_{a}^{i}\right\rangle &=&0\,,  \label{DJ 1}
\\
D^{(0)}_i\left\langle \bar{J}_{a}^{i}\right\rangle &=&-4\kappa
\,g_{ab}\,\varepsilon ^{ij}F^{(0)b}_{ij}\,,  \label{DJ 2}
\end{eqnarray}
where here $D^{(0)}$ and
$F^{(0)a}=dA^a+\frac{1}{2}\,{f_{bc}}^a\,A^b A^c$ are associated to
the boundary field $A$. We conclude that $J_{a}^{i}$ is
covariantly preserved, while $\bar{J}_{a}^{i}$ exhibits a
covariant non-Abelian anomaly \cite{Bardeen-Zumino}.

To end, note that the choice of the boundary term in (\ref{CS
non-Abelian}) can be understood from the point of view of
asymptotic symmetries. It was shown in
\cite{Arcioni-Blau-O'Loughlin} that, for a three-dimensional CS
theory $I\left[ B,\bar{B}\right] =\alpha \,I_{CS}\left[ B\right]
+\beta \,I_{CS}\left[ \bar{B}\right] +\Omega\left[
B,\bar{B}\right] $, the surface integral in (\ref{CS
non-Abelian})\ is the unique boundary term (up to the chirality
$x^{+}\leftrightarrow x^{-}$) quadratic in gauge fields which
preserves the maximal number of asymptotic symmetries $G_{D}$,
and it exists only for $\alpha +\beta =0$. In consequence,
the asymptotic symmetry of
three-dimensional gravity with torsion (which implies $\alpha +\beta \neq 0$)
and gauge group $SO(1,2)\times SO(1,2)$ \cite{Blagojevic-Vasilic}
is always smaller than $SO(1,2)$.
Therefore, in this case the Lorentz
symmetry cannot be preserved asymptotically, resulting in
a Lorentz anomaly in the dual quantum theory.

\section{First order gravitational theories}

The motivation for this section is twofold. On the one hand, we
are interested in studying the holographic renormalization method
for a gravitational theory (Chern-Simons gravity in five
dimensions \cite{Chamseddine}) which is quadratic in the curvature
and has non-trivial torsional degrees of freedom. We would like to
uncover the role of the spin connection in the AdS/CFT
correspondence as the source for the spin current.

A different motivation to carry out this study follows by an
analogy with three-dimensional Chern-Simons theory. It is well
known that flat connections in three dimensions give rise to
conformal structures in two dimensions \cite{Witten89}. The 3d
Chern-Simons equations $g_{ab}\,F^b=0$ are extended to five
dimensions as $g_{abc}\,F^b F^c=0$. In three dimensions on a
manifold $\Re\times M_2$, it is direct to see that $F=0$ projected
to $M_2$ give rise to conformal Ward identities
\cite{Drinfeld-Sokolov} (for recent discussions see
\cite{Banados-Caro,deBoer-Goeree}.) It is a natural question to
ask whether the five-dimensional theory on $\Re\times M_4$ gives
rise to conformal structures on $M_4$.  (See \cite{Losev-Moore-} and \cite{Nair}
for earlier discussions of these issues.)  We shall analyze these
questions in the particular case of Chern-Simons gravity in five
dimensions, and conclude that the correct Ward identities expected
for a Lorentz and diff invariant theory on $M_4$ can in fact be
derived from it.

\subsection{First order gravitational sources and their Ward identities}
\label{FOG}

In this section we consider CFT deformations defined by first
order theories of gravity with non-trivial torsional degrees of
freedom.

To fix the ideas, consider the example of a Dirac spinor field in
four dimensions coupled to an external gravitational field. The
action is
\begin{equation}\label{Dirac}
I_{e,\omega}[\psi] = \int d^4x \  \left\vert e\right\vert\ i\bar
\psi \,\gamma^i \left(\partial _i +\frac{i}{2}\, \omega^{ab}_{\ i}
\,\gamma_{ab}\right) \psi+\mbox{c.c.}\,,
\end{equation}
where $\left\vert e\right\vert = \sqrt{|g|}$ is the determinant of
the vielbein field, and $\omega^{ab}_{\ i}$ is the spin
connection. This action depends on both gravitational fields $e^a$
and $\omega^{ab}$.

In the standard situation of Riemannian geometry, the spin
connection is regarded to be a function of the vielbein,
$\omega=\omega(e)$, determined by solving the torsion equation.
However, from the point of view  of the action (\ref{Dirac}),
these fields are not subject to variations and it is not mandatory
to link them. In fact, if one could compute the effective action
\begin{equation}\label{1}
e^{iW[e,\omega]} = \int D\psi \, e^{iI_{e,\omega}[\psi]},
\end{equation}
for arbitrary values of $e^a$ and $\omega^{ab}$, the functional
$W[e,\omega]$ would carry more information as it would have two
independent arguments.

This is the calculation we attempt to do using holography.
We shall consider a five-dimensional
gravitational action having spin connection degrees of freedom.
This means that its solutions are characterized by independent
values of the (boundary) vielbein and spin connection, and the
on-shell renormalized action is a four-dimensional functional
$I_{ren}[e,\omega]$. Applying the rules of AdS/CFT correspondence
in a classical gravity approximation, we shall identify\footnote{Of
course, we do not claim that the dual of Chern-Simons gravity is a
Dirac field. We have used the action (\ref{Dirac}) only to exhibit
the structure of Ward identities we aim to derive.}
$I_{ren}[e,\omega]=W[e,\omega]$ and prove that the bulk
constraints in the gravitational theory lead to the correct Ward
identities for the 1-point functions derived from $W[e,\omega]$.

More concretely, we are interested in the 1-point functions
\begin{equation}
\tau _{~a}^{i}(x)=\frac{1}{\left\vert e\right\vert }\frac{\delta
W[e,\omega ] }{\delta e_{i}^{a}(x)}\,,\qquad  \sigma
_{~ab}^{i}(x)=\frac{1}{ \left\vert e\right\vert }\frac{\delta
W[e,\omega ]}{\delta \omega _{\ i}^{ab}(x)}\,.
\end{equation}
The tensor $\tau_a$ is related to the energy momentum tensor of
the theory.\footnote{The energy momentum tensor is a Hodge dual to
the 3-form $\tau_a$, and it has the components  $
{\tau^i}_a=\frac{1}{3!\left\vert e\right\vert }\,\,\varepsilon
^{ijkl}\,\tau_{a\,jkl}$ (and similarly for $\sigma_{ab}$).} The
tensor $\sigma_{ab}$ will be called ``spin current". Note that
$\tau_a$ is not equal to the second order energy-momentum tensor
in which $\omega=\omega(e)$.

The Ward identities we expect to find are the following. First
consider the invariance of the action (\ref{1}) under Lorentz
transformations
\begin{eqnarray}
\delta_\lambda e_i^{a} &=&-\lambda^a_{\ b}\,e^b_i\,,  \label{Lorentz-e}\\
\delta_\lambda \omega_{\ i}^{ab} &=&D_i\lambda ^{ab}\,.
\label{Lorentz-w}
\end{eqnarray}
If the measure is also Lorentz invariant then $W$ must be
invariant as well. This implies \begin{eqnarray} 0 &=&\int
d^{4}x\,\left\vert e\right\vert \left( -\lambda ^{ab}\tau
_{~a}^{i}e_{_{i}b}+\frac{1}{2}\,\sigma _{~ab}^i\,D_{i}\lambda ^{ab}\right) \\
&=&\frac{1}{2}\int d^{4}x\,\lambda ^{ab}\left[ 2\left\vert
e\right\vert \tau _{ab}-D_{i}\left( \left\vert e\right\vert \sigma
_{~ab}^{i}\right) \right] \,,
\end{eqnarray}
where $\tau _{ab}=e_{ia}\tau^i_{\ b}$. After antisymmetrization in
$[ab]$, we arrive at the conservation law (Ward identity) for the
Lorentz transformations
\begin{equation}\label{WL}
D_{i}\left(\left\vert e\right\vert\sigma
_{~ab}^i\right)=\left\vert e\right\vert \left( \tau _{ab}-\tau
_{ba}\right)\,.
\end{equation}
When $\sigma$ vanishes, the energy momentum tensor is symmetric,
as expected in the torsionless case.

Next, we consider diffeomorphisms. The action (\ref{Dirac})
coupled to $e$ and $\omega$ is also invariant under
diffeomorphisms\footnote{ This form of diff transformations is
called improved diffeomorphisms \cite{Jackiw} and it differs from
a Lie derivative by a Lorentz transformation with parameter
$\lambda^{ab}=\xi^i\omega_{\ i}^{ab}$.} with parameter $\xi^i$,
\begin{eqnarray}
\delta_\xi e_{i}^{a} &=&D_{i}\left( \xi ^{j}e_{j}^{a}\right) +\xi
^{j}T_{\ ij}^{a}\,,   \\
\delta_\xi \omega _{\ i}^{ab} &=&\xi ^{j}R_{\ \ ij}^{ab}\,.
\label{improved diff}
\end{eqnarray}
Again assuming invariance of the measure, this symmetry gives rise
to the equation
\begin{eqnarray}
0&=&\int d^{d}x\,\xi ^{j}\left[ -D_i\left( \left\vert e\right\vert
\tau _{~a}^{i}\right) e_{j}^{a}+\left\vert e\right\vert \left(
\tau _{~a}^{i}T_{\ ij}^{a}+\frac{1}{2}\,\sigma _{~ab}^{i}R_{\ \
ij}^{ab}\right) \right] \,,
\end{eqnarray}
or, eliminating the vector $\xi^i$, we get
\begin{equation}\label{WD}
D_{i}\left( \left\vert e\right\vert \tau _{~a}^{i}\right)
e_{j}^{a}=\left\vert e\right\vert \left( \tau _{~a}^{i}T_{\
ij}^{a}+\frac{1}{2 }\,\sigma _{~ab}^{i}R_{\ \ ij}^{ab}\right) \,.
\end{equation}
The meaning of this equation is clear after rewriting it in the
form $\left( \left\vert e\right\vert \tau _{~j}^i\right) _{;\,i}=$
$\frac{\left\vert e\right\vert}{2 }\,\sigma _{~ab}^{i}R_{\ \
ij}^{ab}$. When $\sigma$ vanishes, the energy momentum tensor is
covariantly preserved, as known in the torsionless case.

Equations (\ref{WL}) and (\ref{WD}) are the two Ward identities
that we expect to derive by an AdS/CFT interpretation of
five-dimensional Chern-Simons gravity.

\subsection{First order actions}

Our first goal is to find a gravitational action which has spin
connection (or torsional) degrees of freedom. In other words, we
seek a gravitational action whose equations of motion leave
$\omega^{ab}$ as an independent field.

The Einstein-Hilbert action (with or without cosmological term and
in any dimension) does not satisfy this condition because the
torsion vanishes and spin connection is related to the vielbein.
The Palatini formalism does not change this conclusion because one
of the equations of motion of the Einstein-Hilbert action implies,
precisely, that the  torsion is zero.

There exist, however, gravitational actions for which the spin
connection is an independent field. In fact, generically, all
gravitational actions containing quadratic or higher powers of the
curvature tensor will have this property. For actions of the form
$R^n$ ($n>1$ and $R$ represents the scalar, Ricci or Riemann
curvature tensor with indices properly contracted), the equations of motion
that follow by varying the metric and connection independently are
not equivalent to those in which the torsion is assumed to vanish
from the start.  In other words, the Palatini and second order
formalisms are not equivalent for these actions.  To our knowledge,
this phenomenon was first discussed in
\cite{Banados-Garay-Henneaux}, in the context of Chern-Simons
gravities.  Within AdS/CFT, actions with quadratic powers in the
curvature tensor and Ricci scalar have been considered in
\cite{Nojiri-Odintsov-Blau-Narain-Gava,Schwimmer-Theisen}.

The family of Lovelock actions of gravity is of particular
interest because the equations of motion remain of second order
(two derivatives) and that makes the analysis simpler. A
particular case of the  Lovelock action, which simplifies the
analysis even further, is given by the Chern-Simons gravity action
in five dimensions
\begin{equation}\label{Lovelock}
I[e,\omega] =\kappa\int_{M_5}
\varepsilon_{ABCDE}\left(\hat{R}^{AB} \hat{R}^{CD} \hat{e}^E + {2
\over 3} \, \hat{R}^{AB}  \hat{e}^{C} \hat{e}^ D \hat{e}^E + {1
\over 5}\, \hat{e}^A \cdots \hat{e} ^E \right)\,,
\end{equation}
where hatted fields are five-dimensional objects. Here
$\hat{R}^{AB}$ is the curvature two form and the AdS radius $
\ell$ is set to 1. Apart from diffeomorphism invariance, the
action is also invariant under the gauge group AdS$_5$, or
$SO(2,4)$. Varying $I$ with respect to $e^A$ and $\omega^{AB}$, we
obtain the equations of motion
\begin{eqnarray}
  \varepsilon_{ABCDE} \left( \hat{R}^{AB} + \hat{e}^A \hat{e}^B\right)
  \left(\hat{R}^{CD} + \hat{e}^C \hat{e}^D\right) &=& 0\,,  \label{R}\\
  \varepsilon_{ABCDE}\left( \hat{R}^{AB} + \hat{e}^A \hat{e}^B\right)
  \hat{T}^C &=& 0\,. \label{T}
\end{eqnarray}
Equation (\ref{T}), associated to the variation of $\omega^{AB}$,
can be solved by the torsion condition $\hat{T}^A=0$. However,
this is not the most general solution.  In fact, equation
(\ref{T}) possesses a much larger space of solutions. See
\cite{Banados-Garay-Henneaux} for a detailed analysis of its
dynamical structure.

For our purposes here, we only recall that the induced boundary
values of the spin connection, $e^a_i$ and $\omega^{ab}_{\ i}$,
can be taken as fully independent, and will be identified with the
CFT sources discussed above.

The holographic description of dual anomalies for Chern-Simons
gravity with no torsion  has been studied in
\cite{Banados-Schwimmer-Theisen,Banados-Olea-Theisen}.  The role
of gravitational Chern-Simons forms for AdS$_3$/CFT$_2$ has
recently been analyzed in
\cite{Kraus-Larsen-Solodukhin-Babington}, assuming also that the
torsion vanishes.\footnote{``Gravitational Chern-Simons terms" are
Chern-Simon forms for the Lorentz group $SO(D-1,1)$ and they
depend only on the spin connection.  They differ from
``Chern-Simons Gravities" \cite{Achucarro-Townsend,Witten88} which
are Chern-Simons forms for the AdS group $SO(D-1,2)$ and depend on
both the vielbein and spin connection.}

\subsection{Chern-Simons theories and finite FG expansions
\label{CS theories}}

We now use the AdS/CFT prescription. The first goal is to solve
the equations of motion (\ref{R}) and (\ref{T}) order by order in
an asymptotic expansion near the boundary.

This analysis is actually remarkably simple, given the
Chern-Simons structure of these equations.  Recall that the action
(\ref{Lovelock}) belongs to the general set of five-dimensional
Chern-Simons actions  $\int AFF + \cdots$, for the particular case
in which the Lie algebra is chosen as $SO(4,2)$
\cite{Chamseddine,Banados-Troncoso-Zanelli}. The equations of
motion (\ref{R}) and (\ref{T}) can be combined in the
form
\begin{equation}\label{CSg}
g_{abc}\,\varepsilon^{\mu\nu\lambda\rho\sigma} F^b_{\ \mu\nu}\,
F^c_{\ \lambda\rho}=0\,,
\end{equation}
where $F^a$ is the $SO(4,2)$ curvature, and the invariant tensor
$g_{abc}$ is related to the Levi-Cevita form. (The components of the
$SO(4,2)$ curvature $F^a$ are $\hat{R}^{AB} + \hat{e}^A \hat{e}^B$
and $\hat{T}^C$.) Note that the content of this paragraph is valid
for any Lie algebra ${\cal G}$.  Lower case Latin indices
$a,b,c,\ldots$ refer to the corresponding adjoint representation.
This notation should not be confused with that of next section in
which Latin indices refer to 4-dimensional Lorentz indices.

We assume that the manifold has the asymptotic structure $\Re
\times M_4$ and that it is parameterized by the local coordinates
$x^\mu = (r,x^i)$. Here $r $ is related to the FG radial
coordinate and $M_4$ is where the CFT lives. The gauge field
is expanded as $A_\mu=(A_r, A_i)$ and the equations (\ref{CSg})
split as
\begin{eqnarray}
g_{abc}\, \varepsilon^{ijkl} F^b_{\ ij}\, F^c_{\ kl} &=& 0\,, \label{CS1} \\
g_{abc}\, \varepsilon^{ijkl} F^b_{\ ij}\, F^c_{\ kr} &=& 0\,.
\label{CS2}
\end{eqnarray}
Note that (\ref{CS1}) contains only derivatives with respect to
$x^i$ and in that sense it is a ``constraint" that must hold at
all values of $r$. The Ward identities of the holographic CFT are
contained in this set of equations.   Equation (\ref{CS2}) depends
on $F_{i\,r}$ which does contain derivatives with respect to $r$.
The structure of the FG expansion is controlled by this component
of the curvature.

Equations (\ref{CS1}) and (\ref{CS2}) are entangled; the space of
solutions of (\ref{CS2}) depends on the solution of (\ref{CS1}).
To give just one example suppose we solve (\ref{CS1}) by $F^a_{\
ij}=0$. Then, (\ref{CS2}) would imply that $F^a_{\ ir}$ is fully
arbitrary. Of course (\ref{CS1}) has other more interesting
solutions for which $F_{ir}$ becomes constrained. A careful analysis of
this point can be found in \cite{Banados-Garay-Henneaux}.
Conditions ensuring that all equations in (\ref{CS1},\ref{CS2})
are functionally independent, have been discussed in
\cite{Miskovic-Troncoso-Zanelli,Miskovic}.

For our AdS/CFT applications, in which boundary fields play the
role of sources, we need the most general solution of (\ref{CS1}).
For a generic solution of (\ref{CS1}), it is shown in
\cite{Banados-Garay-Henneaux} that (\ref{CS2}) implies
\begin{equation}\label{dy}
F^a_{\ ri} = F^a_{\ ij}\, N^j\,,
\end{equation}
where $N^j$ are arbitrary functions.

Now, recalling that $F^a_{\ ri} = \partial _r\, A^a_i - D_i
A^a_r$, where $D_i$ is the covariant derivative in the connection
$A_i$, equation (\ref{dy}) can be written as
\begin{equation}\label{dy2}
\partial _r\, A^a_i = D_i A^a_r + F^a_{\ ij}\, N^j.
\end{equation}
The role of the functions $A_r$ and $N^i$ should now be clear. The
first term on the r.h.s. of (\ref{dy2}) represents a gauge
transformation, while the second is a (improved) diffeomorphism
with parameter $N^i$. This means that the value of $A_i(r+ \delta
r)$ is obtained from $A_i(r)$ by means of a gauge transformation
plus a diffeomorphism. The functions $A_r$ and $N^i$ are the
corresponding parameters of these transformations, and we can
choose them at will.

The simplest gauge choice $A_r=0$ and $N^i=0$ would led to a
trivial FG expansion in which the fields are independent of $r$.
In the applications to Chern-Simons gravity, however, this is not
allowed because setting $A_r=0$ would lead to a degenerate
vielbein and metric.

The simplest non-degenerate choice is
\begin{equation}\label{gauge}
A_r = \mbox{constant Lie algebra element}\,, \qquad  N^i=0\,.
\end{equation}
Equation (\ref{dy2}) then becomes $\partial _r A_i = [A_i,A_r]$,
whose general solution is
\begin{eqnarray}\label{FGCS}
A_i(r,x^j) = e^{-r A_r}\, A_i(0,x^j)\, e^{r A_r} \,,
\end{eqnarray}
where $A(0,x)$ does not depend on $r$.

The field $A_i(r,x)$ is therefore completely fixed once we give
the initial condition   $A_i(0,x)$. However, we must recall that
the initial conditions are not all arbitrary because they must
satisfy the constraints (\ref{CS1}). These constraints can be
easily translated into equations for $A_i(0,x)$ as follows. Since
the $r$-dependence of $A_i(r,x)$ is a pure gauge transformation,
we find $F_{ij}(r,x)= e^{-r A_r}\, F_{ij}(0,x)\, e^{r A_r}$, where
$F_{ij}(0,x)$ is constructed with $A_i(0,x)$. The constraint on
$A_i(0,x)$ is thus simply
\begin{eqnarray}\label{CCS}
g_{abc}\,\varepsilon^{ijkl} F^b_{\ ij}(0,x)\,F^c_{\ kl}(0,x)=0\,.
\end{eqnarray}
As we shall see in the next subsection, these constraints contain
the Ward identities (\ref{WL}) and (\ref{WD}) discussed above.

The expansion (\ref{FGCS}) can be put in a more explicit form by
going to the Cartan-Weyl basis of the Lie algebra. Let
$H_n,E_\alpha$ be this basis with the root system $\alpha_n$, so
that $[ H_n, E_\alpha ] = \alpha_n\, E_\alpha$. Without loss of
generality, we assume that $A_r$ lies along the Cartan subalgebra,
$A_r=\sum_n c^n H_n$, where the coefficients $c^n$ are constant.
The initial condition has the most general form
$A_i(0,x)=\sum_{n}A^n_i(x)\,H_n +\sum_\alpha
A^\alpha_i(x)\,E_\alpha$. The expression (\ref{FGCS}) can be
computed by means of the identity $e^{cH_n} E_\alpha e^{-cH_n} =
e^{c\alpha_n}E_\alpha$, and we find the expansion
\begin{eqnarray}\label{FG5d}
A_i(r,x) = \sum_n A^n_i(x) H_n + \sum_\alpha e^{-r\sum_n c^n
\alpha_n} A_i^\alpha(x) E_\alpha\,.
\end{eqnarray}
Defining $1/ \rho = e^{2r}$, one obtains a finite FG expansion
with powers determined by the roots $\alpha$. For 2+1 gravity this
analysis was performed in detail in \cite{Banados-Chandia-Ritz}.
An expansion of this type was used in \cite{deBoer-Goeree} in a
derivation of W-algebras from $SL(2,N)$ Chern-Simons theories.  We
compute (\ref{FG5d}) for the particular case of 5d Chern-Simons
gravity in the next paragraph.

We end this section with two comments. First, note that even
though our FG expansions are finite, the solutions cannot be
extended to the whole manifold.  The reason is that the gauge
choice (\ref{gauge}) may not be extendible beyond the asymptotic
region. Several patches may be necessary and this depends on
global properties on the manifold and Lie group under
consideration. In this manuscript, we restrict ourselves to the
asymptotic analysis and relevant Ward identities.  The global
analysis is important to find the non-local dependence of the vevs
(subleading terms in a FG expansion) on the sources. We hope to came
back to this problem elsewhere.

Second, note that the solutions considered here, defined by the
expansion (\ref{FGCS}), cannot be seen as fluctuations of AdS
space, $F_{\mu\nu}=0$. Our solutions have $F_{ri}=0$, but $F_{ij}$
must be different from zero. This component of the curvature
cannot be deformed continuously to zero keeping the expansion
(\ref{FGCS}) valid. As we mentioned before, $F_{ij}=0$ is a
degenerate solution of (\ref{CS1}) that trivializes (\ref{CS2}).

\subsection{Explicit FG structure of Chern-Simons gravity}

We now go back to the particular case of Chern-Simons Gravity in
five dimensions, for which the Lie algebra is $SO(4,2)$, and the
gauge field is expanded in terms of the $SO(4,2)$ generators
$P_A\equiv J_{A6},\,J_{AB}$ as ($A,B=1,\dots,5$)
\begin{equation}\label{ASO}
A_\mu  = \hat e^A_{\mu} P_A + {1 \over 2} \, \hat \omega^{AB}_{\ \
\mu} J_{AB}\,.
\end{equation}
The fields $\hat \omega^{AB}$ and $\hat e^A$  satisfy the set of
five-dimensional equations (\ref{R}) and (\ref{T}). As we have
mentioned before, these equations are of the form $F \wedge F=0$,
and all the results of last paragraph apply to this case.

We start by imposing the analogues of (\ref{gauge}) for this
particular Lie algebra.  This means fixing $\hat e^A_r$ and $\hat
\omega^{AB}_{\ \ r}$.  Splitting the Lorentz indices as $A=(a,5)$,
we set $A_r=P_5$ or
\begin{equation}\label{5dgauge}
 \hat e^5_r = 1\,, \qquad \hat e^a_r=0\,, \qquad \hat \omega^{AB}_{\ \ r}=0\,.
\end{equation}
Having fixed the radial components of the gauge field, the radial
dependence of the tangent components $\hat e^A_i(r,x)$ and
$\hat\omega^{AB}_{\ \ i}(r,x)$ becomes completely determined by
(\ref{FGCS}). Next, we shall impose one extra condition on the
boundary vielbein, namely,
\begin{equation}\label{ei=0}
 \hat e^5_i=0\,.
\end{equation}
This condition breaks the five-dimensional Lorentz symmetry down
to a four-dimensional one. It also leaves a four-dimensional
tetrad as a gravitational source.

The calculation of (\ref{FGCS}) for the $SO(4,2)$ algebra is a
simple exercise. In four-dimensional notation the SO(4,2)
generators $P_{A},J_{AB}$ are split as $P_5$, $P_a$, $J_{a5}$,
$J_{ab}$. It will be convenient to define the combinations
\footnote{The non-zero commutators of the full $SO(4,2)$ algebra
are
\begin{equation}
\begin{array}{ll}
\left[ {J}_{ab},{J}_{cd}\right] =\eta _{ad}\,\,{J} _{bc}-\eta
_{bd}\,{J}_{ac}-\eta _{ac}\,{J}_{bd}+\eta _{bc}\, {J}_{ad}\,, &
\left[ {J}_{a}^{+},{J}_{b}^{-}\right]
=2\,\left( -\eta _{ab}\,{P}_{5}+{J}_{ab}\right) \,, \\
\left[ {J}_{ab},{J}_{c}^{\pm }\right] =\eta _{bc}\,{J} _{a}^{\pm
}-\eta _{ac}\,{J}_{b}^{\pm }\,, & \left[ {J} _{a}^{\pm
},{P}_{5}\right] =\pm {J}_{a}^{\pm }\,.
\end{array}
\end{equation}}
\begin{equation}\label{Jpm}
{J}_{a}^{\pm }\equiv {P}_{a}\pm {J}_{a5}.
\end{equation}
Expanding the initial condition ${A} \left( 0,x\right) $ in
(\ref{FGCS}) as
\begin{equation}
{A}\left( 0,x\right) =e^{a}\left( x\right) \,{J}
_{a}^{+}+k^{a}\left( x\right) \,{J}_{a}^{-}+\frac{1}{2}\,\omega
^{ab}\left( x\right) \,{J}_{ab} \label{A(0)}
\end{equation}
(from now on all forms are four-dimensional) and using the
identity $e^{-\alpha P_5}J^\pm_a\, e^{\alpha P_5}$ $=$
$e^{\pm\alpha}J^\pm_a $, we obtain
\begin{equation}
{A}\left( \rho ,x\right) =\frac{1}{\sqrt{\rho }}\,e^{a}\,{J}
_{a}^{+}+\sqrt{\rho }\,k^{a}\left( x\right)
\,{J}_{a}^{-}+\frac{1}{2} \,\omega ^{ab}\left( x\right)
\,{J}_{ab}\,,\label{A(r)}
\end{equation}
where we have returned to the usual FG radial coordinate
$1/\rho=e^{2r}$.

From (\ref{ASO}) and (\ref{Jpm}) we can also write the explicit
relationship between the five-dimensional fields $\hat e^A,\hat
\omega^{AB}$ and the components of (\ref{A(0)}),
\begin{eqnarray}
\hat{e}^{a}\left( \rho ,x\right) &=&\frac{1}{\sqrt{\rho }}\left[
e^{a}(x)+\rho \,k^{a}(x)\right] \,,   \\
\hat{\omega}^{a5}\left( \rho ,x\right) &=&\frac{1}{\sqrt{\rho
}}\left[
e^{a}(x)-\rho \,k^{a}(x)\right] \,,  \label{FG expansion} \\
\hat{\omega}^{ab}\left( \rho ,x\right) &=&\omega ^{ab}(x)\,.
\end{eqnarray}
These expansions can be seen as a finite FG expansion. The
five-dimensional line element $\hat{e}_{\mu}^{\ A}\,\hat{e}_{A\nu
}\,dx^\mu dx^\nu $ has the finite FG form
\begin{equation}\label{FGk}
ds^2 = {d\rho^2 \over 4\rho^2 }  + {1 \over \rho}\left[g_{ij} +
\rho(k_{ij}+ k_{ji}) + \rho^2{k_i}^l k_{lj}\right] dx^i dx^j\,,
\end{equation}
where $g_{ij}={e_i}^a\, e_{ja}$ and $k_{ij} = e_{ia}\,{k_j}^a$.

So far we have solved the equations (\ref{CS2}) for the particular
case of the $SO(4,2)$ Lie algebra. We now write the constraints
(\ref{CS1}) in terms of the basic fields $e^a,k^a,\omega^{ab}$. To
this end, it is useful to have $\hat T^A$ and $F^{AB}=\hat
R^{AB}+\hat e^A \hat e^B$ in terms of the fields $e,\,k,\,\omega$,
\begin{equation}
\begin{array}{ll}
\hat{T}^{5}\left( \rho \right) =-2\,e^{a}k_{a}\,,\smallskip &
F^{a5}\left( \rho \right) =\frac{1}{\sqrt{\rho }}\left( T^{a}-\rho
\,Dk^{a}\right) \,,
\\
\hat{T}^{a}\left( \rho \right) =\frac{1}{\sqrt{\rho }}\left(
T^{a}+\rho \,Dk^{a}\right) \,,\qquad & F^{ab}\left( \rho \right)
=R^{ab}+2\left( e^{a}k^{b}-e^{b}k^{a}\right) \,,
\end{array}
\label{expansion of F}
\end{equation}
where $R^{ab}$ and $T^{a}$ are four-dimensional curvature and torsion,
respectively, and $D=D(\omega )$. Note that these are also finite series in $\rho$.

Inserting these expansions into the constraint equations
(\ref{CS1}), namely, the components of (\ref{R}) and (\ref{T})
tangent to the boundary, we obtain
\begin{eqnarray}
C &\equiv &\varepsilon _{abcd}\,F^{ab}F^{cd}=0\,,  \label{con1} \\
C_{a} &\equiv &\varepsilon _{abcd}\,F^{bc}\,T^{d}=0\,,
\label{con2}
\\
\bar{C}_{a} &\equiv &\varepsilon _{abcd}\,F^{bc}Dk^{d}=0\,,
\label{con3} \\
C_{ab} &\equiv &\varepsilon _{abcd}\,\left(
F^{cd}\,e^{e}k_{e}+2T^{c}Dk^{d}\right) =0\,,  \label{con4}
\end{eqnarray}
where $F^{ab}=R^{ab}+2e^{a}k^{b}-2e^{b}k^{a}$.

This set of $1+4+4+6=15$ equations gives relations between the
fields $e^a,k^a$ and $\omega^{ab}$. We would like to argue now
that (\ref{con1}-\ref{con4}) leave the fields $e^a_i$
and $\omega^{ab}_{i}$ arbitrary. They can then be interpreted as sources.
Furthermore, the field $k$ will be associated to the vevs or
1-point functions, and equations (\ref{con1}-\ref{con4}) imply the
correct Ward identities for them.

\subsection{1-point functions and Ward identities}
\label{Ward}

For orientation, it is useful at this point to recall the standard
gravitational analysis of holographic renormalization
\cite{Skenderis,Henningson-Skenderis,Haro-Solodukhin-Skenderis},
and compare it with our analysis. For the Einstein-Hilbert action,
the asymptotic solution has the expansion
\begin{equation}\label{FGEE2}
ds^2 = {d\rho^2 \over 4\rho^2} + {1 \over \rho} \left( g_{(0) ij}
+  \rho\, g_{(1) ij} + \rho^2\, (g_{(2) ij} + \log\rho\,
h_{(2)ij}) + \cdots \right)dx^idx^j,
\end{equation}
where the coefficients are determined by the Einstein equations.
In a 4+1 decomposition, with $\rho$ playing the role of ``time",
these equations are separated into 10 ``dynamical" (radial
derivatives) equations plus 5 constraints. The 10 dynamical
Einstein equations are of second order in the radial derivative
and they require two initial conditions for their integration. In
fact, plugging the series (\ref{FGEE2}) into the dynamical
equations one finds that $g_{(0)ij}$ and $g_{(2)ij}$ are left
completely arbitrary while all other coefficients in the expansion
(\ref{FGEE2}) are fixed in terms of them. $g_{(0)ij}$ and
$g_{(2)ij}$ are in that sense ``conjugate" variables and, in fact,
the variation of the (renormalized) action can be written as
$\delta I_{ren} = 2\int \tau^{ij} \delta g_{(0)ij}$, where
$\tau^{ij}$ is linear in $g_{(2)ij}$.  The tensor $\tau^{ij}$ is
identified with the holographic energy momentum tensor.  Now,
besides the dynamical equations, there are also 5 constraints
which impose conditions on $g_{(2)ij}$, or $\tau^{ij}$. These
constraints turn out to be exactly the Ward identities on the
trace and divergence of $\tau^{ij}$
\cite{Skenderis,Henningson-Skenderis}. (See \cite{Corley} for a
generic analysis of these relations.)

Our analysis follows a similar structure but with more fields. The
initial conditions $A_i(0,x)$ appearing in (\ref{FGCS}) are the
analogue of $g_{(0)ij}$ and $g_{(2)ij}$.   The ``dynamical"
equations (\ref{CS2}) impose no conditions on them.  The
constraints (\ref{CS1}), on the other hand, impose conditions on
$A_i(0,x)$, which should be related to the holographic Ward
identities.   For the particular Lie algebra $SO(4,2)$, these
constraints are given by equations (\ref{con1}-\ref{con4}), where
$A_i(0,x)$ has been expanded in terms of $e^a_i,{k_i}^a$ and
$\omega^{ab}_{\ i}$ in (\ref{A(0)}).

The rest of the program proceeds just like for standard gravity,
with minor variations. We shall  prove that the variation of the
renormalized bulk action can be expressed as ($M_4=\partial M_5$)
\begin{equation}\label{dIren}
\delta I_{ren} = \int_{M_4} (\delta e^a \,\tau_a +\frac{1}{2}\,
\delta \omega^{ab}\,\sigma_{ab} ).
\end{equation}
The coefficients $\tau_a$ and $\sigma_{ab}$ both depend on $k$ and
will be interpreted as the holographic energy momentum tensor and
spin current. Finally, we show that the constraints
(\ref{con1}-\ref{con4}) imply the Ward identities (\ref{WL}) and
(\ref{WD}) for these tensors, in full analogy with the pure
gravity case.

We start with the action (\ref{Lovelock}) which yields the
equations (\ref{R}) and (\ref{T}).  This action will need boundary
terms to be well defined for fixed $e^a$ and $\omega^{ab}$, and
will also need to be renormalized to make it finite.

Varying (\ref{Lovelock}) on-shell and keeping all boundary terms
on-shell, one finds
\begin{equation}
\delta I=-2\kappa \int\limits_{M_{4}}\varepsilon
_{ABCDE} \left( \hat{R}^{AB}+\frac{1}{3}\,\hat{e}^{A}\hat{e}^{B}\right)
\hat{e}^{C}\delta \hat{\omega}^{DE}.  \label{variation}
\end{equation}

Next, we evaluate this variation on the asymptotic solution
described above. By a direct calculation this can be written as
\begin{equation}
\delta I=4\kappa
\int\limits_{M_{4}}\varepsilon _{abcd}\,\left[ \left(
R^{ab}+2\,e^{a}k^{b}\right) \left( -k^{c}\delta e^{d}+e^{c}\delta
k^{d}\right) +\left( De^{a}k^{b}-Dk^{a}e^{b}\right) \delta \omega
^{cd} \right] +\delta V\,,  \label{varIren}
\end{equation}
where
\begin{equation}
V= \frac{2\kappa }{3\rho^2 }\int\limits_{M_{4}}\varepsilon
_{abcd}e^{a}e^{b}e^{c}e^{d} -\frac{2\kappa }{\rho
}\int\limits_{M_{4}}\varepsilon _{abcd}\left( R^{ab}+{4\over
3}\,e^{a}k^{b}\right) k^{c}e^{d}\,.
\end{equation}
(We have omitted terms with positive powers of $\rho$ because they
cancel in the limit $\rho\rightarrow 0$.)

Note that $V$ is divergent as $\rho\rightarrow 0$,  while all
other terms in (\ref{varIren}) are finite (and
$\rho$-independent). The divergencies thus appear in the form of a
total variation which can be subtracted from $I$. We define the
renormalized action $I^{'}_{ren} = I-V$ whose variation is given
by the first piece in (\ref{varIren}). This completes the
``renormalization" problem. But $I_{ren}'$ is not yet the correct
action because we have to deal with the ``Dirichlet problem",
namely, we need to make sure that the action has well defined
variations for $e^a$ and $\omega^{ab}$ fixed.

Having eliminated $V$ in (\ref{varIren}), the remaining terms (in
square bracket) contain variations of $e^a,k^a$ and $\omega^{ab}$.
The variations of $k^a$ can be transformed into variations of
$e^a$ and $\omega^{ab}$ by adding finite boundary terms,
\begin{eqnarray}
4\kappa \,\varepsilon _{abcd}\left( R^{ab}+2\,e^{a}k^{b}\right) e^{c}\delta
k^{d} &=&4\kappa \,\varepsilon _{abcd}\left[ -\left(
R^{ab}+2\,e^{a}k^{b}\right) k^{c}\delta e^{d}+D\left( e^{a}k^{b}\right)
\delta \omega ^{cd}\right] \nonumber\\
&& +\, \delta(\mbox{boundary term})\,,
\end{eqnarray}
where a total derivative has been omitted since it will vanish
under the integral. Discarding all total variations, we finally arrive
at
\begin{equation}
\delta I_{ren} = -8\kappa \int\limits_{M_{4}}\varepsilon _{abcd}\,\left[ \left(
R^{ab}+2\,e^{a}k^{b}\right) k^{c}\delta e^{d}-T^{a}k^{b}\delta
\omega ^{cd} \right] \,,  \label{CFT}
\end{equation}
from where we read off the holographic energy-momentum tensor and
spin current,
\begin{eqnarray}
\tau _{a} &=&-8\kappa \,\varepsilon _{abcd}\left(
R^{bc}+2\,e^{b}k^{c}\right) k^{d}\,,  \label{tau} \\
\sigma _{ab} &=&-16\kappa \,\varepsilon _{abcd}\,T^{c}k^{d}\,.
\label{sigma}
\end{eqnarray}
(We omit writing expectation values $\left\langle \cdots \right\rangle $.)
In components, these tensors are
\begin{eqnarray}
\tau _{~a}^{i} &=&-\frac{8\kappa }{\left\vert e\right\vert
}\,\varepsilon ^{ijkl}\varepsilon _{abcd}\left( \frac{1}{2}\,R_{\
\ jk}^{bc}+2\,e_{j}^{b}
\,k_{k}\,^{c}\right) k_{l}^{\;\,d}\,,  \label{tensor 1} \\
\sigma _{~ab}^{i} &=&-\frac{8\kappa }{\left\vert e\right\vert
}\,\varepsilon ^{ijkl}\varepsilon _{abcd}\,T_{\
jk}^{c}k_{l}^{\;\,d}\,.  \label{tensor 2}
\end{eqnarray}

Our last step is to show that the tensors $\tau _{~a}^{i}$ and
$\sigma _{~ab}^{i}$ satisfy the conservation laws, or Ward
identities (\ref{WL}, \ref{WD}),  associated to Lorentz
transformations and diffeomorphisms.  To prove these identities we
need to use the constraint equations (\ref{con1}-\ref{con4}).

To this end we consider the contraction operator $I_{i}$ which
maps $p$-forms to ($p-1$)-forms.\footnote{ The contraction
operator $I_{i}$ acts on the $p$-form $\Omega =\frac{1}{p!}
\,\Omega _{i_{1}\cdots i_{p}}\,dx^{i_{1}}\cdots dx^{i_{p}}$ as
$I_{i}\Omega = \frac{1}{\left( p-1\right) !}\,\Omega
_{ii_{2}\cdots i_{p}}\,dx^{i_{2}}\cdots dx^{i_{p}}$.} Since any
5-form vanishes on $M_{4}$ and the contraction operator $ I_{i}$
obeys the Leibnitz rule, from $I_{i}(\varepsilon _{abcd}$ $
e^{a}k^{b}k^{c}T^{d})\equiv 0$ and $I_{i}\left( \varepsilon
_{abcd}\,R^{ab}k^{c}T^{d}\right) \equiv 0$, we obtain the
identities
\begin{eqnarray}
\varepsilon _{abcd}\,I_{i}e^{a}k^{b}k^{c}T^{d} &=&\varepsilon
_{abcd}\,\left(
2e^{a}I_{i}k^{b}k^{c}T^{d}+e^{a}k^{b}k^{c}I_{i}T^{d}\right)
\,, \\
\varepsilon _{abcd}\,I_{i}R^{ab}k^{c}T^{d} &=&-\varepsilon
_{abcd}\,R^{ab}I_{i}k^{c}T^{d}+\varepsilon _{abcd}\,R^{ab}k^{c}I_{i}T^{d}\,.
\end{eqnarray}
With the help of these identities and the expressions
(\ref{con2}-\ref{con4}), it can be shown that
\begin{eqnarray}
D\sigma _{ab}-\left( e_{a}\tau _{b}-e_{b}\tau _{a}\right) &=&-
8\kappa\,C_{ab}\,, \\
D\tau _{a}-\left( I_{a}T^{b}\tau
_{b}+\frac{1}{2}\,I_{a}R^{bc}\sigma _{bc}\right) &=&-8\kappa
\left( \bar{C}_{a}-k_{a}^{\;\;b}C_{b} \right) \,.
\end{eqnarray}
Since all constraints vanish on $M_{4}$, the conservation laws
(\ref{WL}, \ref{WD}) are in fact satisfied.

\subsection{Weyl anomaly}

The energy-momentum tensor is also a generator of conformal
transformations, where the conservation law requires that
$\tau^i_{\ a}$ be traceless. With the help of eq. (\ref{con1}),
one can show that
\begin{equation}
e^{a}\tau _{a}=\kappa \,\varepsilon _{abcd}\,R^{ab}R^{cd}-\kappa \,C\,.
\end{equation}
The l.h.s. is the trace of the tensor $\tau _{~a}^{i}$, and since
$C=0$, we obtain that its trace is
\begin{equation}
\tau _{~a}^{a}=\frac{\kappa }{4}\,\varepsilon ^{ijkl}\,\varepsilon
_{nmpq}\,R_{\ \ \ ij}^{nm}\,R_{\ \  kl}^{pq}=\kappa E_{4}\,.
\label{Weyl}
\end{equation}
We reproduce the result that the holographic Weyl anomaly is given
by the Euler density, $E_{4}$
\cite{Henningson-Skenderis,Schwimmer-Theisen,Deser-Schwimmer,Imbimbo-Schwimmer-Theisen-Yankielowicz}.
The torsion does not enter the conformal anomaly explicitly, but
only through the spin connection $\omega^{ab}_{\ i}$ in the
curvature. This is not surprising since there is no
four-dimensional topological invariant with even parity
constructed from the torsion tensor $T^a_{\ i}$. As in
torsion-less Chern-Simons gravity, only Type A anomaly emerges
\cite{Banados-Schwimmer-Theisen,Banados-Olea-Theisen}.

\section{Discussion}

We would like to end with some comments on chiral anomalies for
Chern-Simons gravity duals.  In \cite{Witten98}, Witten noted that
the Chern-Simons term added to Type IIB supergravity on
AdS$_{5}\times S^{5}$ is responsible for the occurrence of the
chiral anomaly in the corresponding CFT. It has been also known
for some time that the fully antisymmetric part of the torsion is
related to the chiral anomaly of spinors in Riemann-Cartan spaces
\cite{Obukhov,Urrutia-Vergara}. These facts suggest that the
Chern-Simons gravity dual theory should exhibit a chiral anomaly,
due to the coupling of the spin connection. The chiral anomaly on
a four-dimensional Riemannian manifold is proportional to another
topological invariant quadratic in curvature, the Pontryagin
density $P_4=R^{ab}R_{ab}$. The torsion on the other hand is seen as
the field strength associated to the vielbein, and it is natural to
ask whether a field theory coupled to both curvature and torsion
would develop an anomaly which depends explicitly on torsion.
A natural candidate to represent this generalized anomaly is the Pontryagin
density for AdS$_4$
\begin{equation}
P_{4}=R^{ab}R_{ab}+\frac{2}{\ell ^{2}}\left(
R^{ab}e_{a}e_{b}-T^{a}T_{a}\right) \,,  \label{Pontryagin R+T}
\end{equation}
where $\ell$ is the AdS radius. When $T^{a}$ vanishes, the term in
parentheses also vanishes, since it can be locally written as
$-d(T^{a}e_{a})$. There is a controversy whether the second
term in (\ref{Pontryagin R+T}), which is by itself a topological
invariant, should contribute to the chiral anomaly or not. See
\cite{Chandia-Zanelli97} and \cite{Kreimer-Mielke} for details.
The AdS/CFT correspondence offers a rich ground to test the
dependence of the chiral anomaly on torsion.

The explicit calculation could be performed as follows. Taking as
a working example the Dirac field (\ref{Dirac}), we see that the
chiral current is proportional to the fully antisymmetric part of
the spin current,
\begin{equation}\label{J-chiral}
J^i_{ch}=\frac{1}{3!}\, \varepsilon^{abcd}\, e^i_a \sigma_{bcd}\,.
\end{equation}
In fact, the spin current for the Dirac field is $ \sigma
^{i}\,_{ab}=\frac{1}{|e|}\frac{\delta I_{D}}{ \delta \omega _{\
i}^{ab}}=i\,\varepsilon _{abcd}\,e^{ic}\bar{\psi} \gamma
_{5}\gamma ^{d}\psi$, and its fully antisymmetric part yields
$J_{ch}^{a}= i\,\bar{\psi}\gamma _{5}\gamma ^{a}\psi \,$, as
expected for the chiral current.

In our case, from (\ref{tensor 2}), we obtain the chiral anomaly
(defined as $\partial_i (|e| J^i_{ch})={|e|\,\cal A}_{ch}$)
\begin{equation} \mathcal{A}_{ch}=\frac{8\kappa
}{3\,|e|}\,\varepsilon ^{jknm}\partial _{i}\left(k_{kj}T_{\
nm}^{i}-k_{k}^{\ i}T_{jnm}\right) \,, \label{A5}
\end{equation}
where $k_{ij}$ is a solution of the field equations
(\ref{con1}--\ref{con4}). The question is whether this expression
coincides with (\ref{Pontryagin R+T}), or its Lorentz
version $R^{ab}R_{ab}$, will be left open. This requires the
general solution (or at least its local part) of $k_{ij}$ which
has escaped us.

Finally, it is worth noticing that the whole analysis of
holographic CFT associated to Chern-Simons gravities can be easily
generalized to any odd dimension, for any extension (including
supersymmetric ones) of the AdS group, since only the gravitational
part requires non-trivial gauge fixing. In all cases, the
Fefferman-Graham expansion is finite or truncated. In the
gravitational sector, the only sources are $e$ and $\omega$
because the undetermined (on-shell) components of $k$ do not
couple to the CFT gravitational current, corresponding therefore
to the Fefferman-Graham ambiguity.

\acknowledgments

We thank Adam Schwimmer for comments on the manuscript. O.M.
thanks Jorge Zanelli for helpful discussions. This work is funded
in parts by grants FONDECYT 1020832, 1060648, 3040026 and 7020832.
O.M. is also supported by PUCV through the program Investigador
Joven 2006. S.T. thanks the Centro de Estudios Cient\'ificos in
Valdivia, where this project was initiated, for hospitality.


\begin{thebibliography}{99}



\bibitem{Maldacena}
J. M. Maldacena, The large $N$ limit of superconformal field
theories, \emph{Adv. Theor. Math. Phys.} \textbf{2} (1998) 231;
\emph{Int. J. Theor. Phys.} \textbf{38} (1999) 1113
[\hepth{9711200}].



\bibitem{Gubser-Klebanov-Polyakov}
S. S. Gubser, I. R. Klebanov and A. M. Polyakov, A semiclassical
limit of the gauge string correspondence, \emph{Nucl. Phys.}
\textbf{B636} (2002) 99 [\hepth{0204051}].



\bibitem{Witten98} E. Witten,  Anti-de Sitter space and
holography, \emph{Adv. Theor. Math. Phys.} \textbf{2} (1998) 253
[\hepth{9802150}].



\bibitem{Aharony-Gubser-Maldacena-Ooguri-Oz}
O. Aharony, S. S. Gubser, J. M. Maldacena, H. Ooguri and Y. Oz,
 Large $N$ Field Theories, String Theory and
Gravity, \emph{Phys. Rept.} \textbf{323} (2000) 183
[\hepth{9905111}].




\bibitem{Skenderis} K. Skenderis,
 Lecture notes on holographic
renormalization, \emph{Class. Quant. Grav.} \textbf{19} (2002)
5849 [\hepth{0209067}].




\bibitem{Henningson-Skenderis} M. Henningson and K. Skenderis,
 The holographic Weyl anomaly,
\emph{JHEP } 9807 (1998) 023 [\hepth{9806087}].




\bibitem{Banados-Garay-Henneaux} M. Ba\~{n}ados, L. J. Garay and M.
Henneaux,  The local degrees of freedom of higher-dimensional pure
Chern-Simons theories, \emph{ Phys. Rev.} \textbf{D53} (1996) 593
[\hepth{9506187}]; M. Ba\~{n}ados, L. J. Garay and M. Henneaux,
The dynamical structure of higher-dimensional Chern-Simons theory,
\emph{Nucl. Phys.} \textbf{B476} (1996) 611 [\hepth{9605159}].



\bibitem{Fefferman-Graham} C. Fefferman and R. Graham,
Conformal invariants, \emph{The mathematical heritage of Elie
cartan }(Lyon 1984), Ast\'{e}risque, 1985, Numero Hors Serie, 95.




\bibitem{Skenderis-Solodukhin} K. Skenderis and S. N. Solodukhin,
 Quantum effective action from the AdS/CFT
correspondence, \emph{Phys. Lett.} \textbf{B472} (2000) 316
[\hepth{9910023}].




\bibitem{Banados-Chandia-Ritz}
M. Ba\~{n}ados, O. Chand\'{i}a and A. Ritz, Holography and the
Polyakov action, \emph{Phys. Rev.} \textbf{D65} (2002) 126008
[\hepth{0203021}].




\bibitem{Banados-Caro} M. Ba\~{n}ados and R. Caro,  Holographic
Ward identities: examples from $2+1$ gravity, \emph{JHEP}
\textbf{0412} (2004) 036 [\hepth{0411060}].




\bibitem{Brown-Henneaux} J.D. Brown and M. Henneaux,  Central
charges in the canonical realization of asymptotic symmetries: an
example from three-dimensional gravity, \emph{Commun. Math. Phys.}
\textbf{104} (1986) 207.




\bibitem{Achucarro-Townsend} A. Ach\'{u}carro and P. K. Townsend,
 A Chern-Simons action for three-dimensional
anti-de Sitter supergravity theories, \emph{Phys. Lett.}
\textbf{B180} (1986) 89.




\bibitem{Witten88} E. Witten,  2+1 dimensional gravity as
an exactly soluble system, \emph{Nucl. Phys.} \textbf{B311 }
(1998) 46.




\bibitem{Witten89} E. Witten,  Quantum field theory
and the Jones polynomial, \emph{Commun. Math. Phys.} \textbf{121}
(1989) 351.




\bibitem{Coussaert-Henneaux-vanDriel}
O. Coussaert, M. Henneaux and P. van Driel,  The asymptotic
dynamics of three-dimensional Einstein gravity with a negative
cosmological constant, \emph{Class. Quant. Grav.} \textbf{12}
(1995) 2961 [\grqc{9506019}].




\bibitem{Banados-Bautier-Coussaert-Henneaux-Ortiz} M. Ba\~{n}ados,
K. Bautier, O. Coussaert, M. Henneaux and M. Ortiz,
 Anti-de Sitter /CFT correspondence in
three-dimensional supergravity, \emph{Phys. Rev.} \textbf{D58}
(1998) 085020 [\hepth{9805165}].




\bibitem{Carlip} S. Carlip,  Quantum gravity in $2+1$
dimensions (Cambridge University Press, 1998).


\bibitem{Arcioni-Blau-O'Loughlin} G. Arcioni, M. Blau and M. O'Loughlin,
 On the boundary dynamics of Chern-Simons
gravity, \emph{JHEP} \textbf{0301} (2003) 067 [\hepth{0210089}].


\bibitem{Bardeen-Zumino}  W. A. Bardeen and B. Zumino,
Consistent and covariant anomalies in gauge gravitational
theories, \emph{Nucl. Phys.} \textbf{B244} (1984) 421.


\bibitem{Blagojevic-Vasilic} M. Blagojevi\'{c} and M. Vasili\'{c},
 3-D gravity with torsion as a Chern-Simons gauge
theory, \emph{Phys. Rev.} \textbf{D68}: 104023 (2003)
[\grqc{0307078}].




\bibitem{Chamseddine} H. Chamseddine,  Topological gauge
theory of gravity in five dimensions and all odd dimensions ,
\emph{Phys. Lett.} \textbf{B233} (1989) 291;  Topological gravity
ans supergravity in various dimensions, \emph{ Nucl. Phys.}
\textbf{B346} (1990) 213.



\bibitem{Drinfeld-Sokolov} V. G.
Drinfeld and V. V. Sokolov,  Lie algebras and equations of
Korteweg-de Vries type, \emph{J. Sov. Math.} \textbf{30} (1984)
1975; \emph{Sov. Math. Doklady} \textbf{3} (1981) 457.



\bibitem{deBoer-Goeree} J. de Boer and J. Goeree,
W Gravity from Chern-Simons theory, \emph{Nucl. Phys.}
\textbf{B381} (1992) 329 [\hepth{9112060}].




\bibitem{Losev-Moore-}
  A. Losev, G. W. Moore, N. Nekrasov and S. Shatashvili,
  Four-dimensional avatars of two-dimensional RCFT,
  \emph{Nucl. Phys. Proc. Suppl.}  {\bf 46} (1996) 130
  [\hepth{9509151}].


 \bibitem{Nair}
  V. P. Nair and J. Schiff,
  Kahler Chern-Simons theory and symmetries of antiselfdual gauge fields,
  \emph{Nucl. Phys.} {\bf B371} (1992) 329.



\bibitem{Jackiw} R. Jackiw,  Gauge covariant
conformal transformations, \emph{Phys. Rev. Lett.} \textbf{41}
(1987) 1635.



\bibitem{Nojiri-Odintsov-Blau-Narain-Gava} S. Nojiri and S. D.
Odintsov,  On the conformal anomaly from higher derivative gravity
in AdS/CFT correspondence, \emph{Int. J. Mod. Phys.} \textbf{A15}
(2000) 413 [\hepth{9903033}]; M. Blau, K.S. Narain and E. Gava, On
subleading contributions to the AdS/CFT trace anomaly, \emph{JHEP}
\textbf{9909} (1999) 018 [\hepth{9904179}].




\bibitem{Schwimmer-Theisen} A. Schwimmer and S. Theisen,
Universal features of holographic anomalies, \emph{JHEP}
\textbf{0310} (2003) 001 [\hepth{0309064}]; Diffeomorphisms,
anomalies and the Feffernam-Graham ambiguity, \emph{JHEP}
\textbf{0008} (2000) 032 [\hepth{0008082}].




\bibitem{Banados-Schwimmer-Theisen} M. Ba\~{n}ados, A. Schwimmer and S.
Theisen,  Chern-Simons gravity and holographic anomalies,
\emph{JHEP} \textbf{0405} (2004) 039 [\hepth{0404245}].




\bibitem{Banados-Olea-Theisen} M. Ba\~{n}ados, R. Olea and S. Theisen,
 Counterterms and dual holographic anomalies in
CS gravity, \emph{JHEP} \textbf{0510} (2005) 067
[\hepth{0509179}].




\bibitem{Kraus-Larsen-Solodukhin-Babington} P. Kraus and F. Larsen,
 Holographic gravitational anomalies  [\hepth{0508218}]; S. N. Solodukhin,
 Holography with gravitational Chern-Simons  [\hepth{0509148}];  S. N. Solodukhin,  Holographic
description of gravitational anomalies [\hepth{0512216}]; J.
Babington, Towards a holographic dual of SQCD: holographic
anomalies and higher derivative gravity [\hepth{0512029}].




\bibitem{Banados-Troncoso-Zanelli}
M. Ba\~{n}ados, R. Troncoso and J. Zanelli, Higher dimensional
Chern-Simons supergravity, \emph{Phys. Rev.}\textbf{ D54} (1996)
2605 [\grqc{9601003}]; R. Troncoso and J. Zanelli,  New gauge
supergravity in seven dimensions and eleven dimensions,
\emph{Phys. Rev.} \textbf{D58} (1998) 101703 [\hepth{9710180}].




\bibitem{Miskovic-Troncoso-Zanelli} O. Mi\v{s}kovi\'{c} and J. Zanelli,
 Dynamical structure of irregular constrained
systems, \emph{J. Math. Phys.} \textbf{44} (2003) 3876
[\hepth{0302033}]; O. Mi\v{s}kovi\'{c}, R. Troncoso and J.
Zanelli,  Canonical sectors of five-dimensional Chern-Simons
theories, \emph{Phys. Lett.} \textbf{B615} (2005) 277
[\hepth{0504055}].



\bibitem{Miskovic} O. Mi\v{s}kovi\'{c},
 Dynamics of Wess-Zumino-Witten and Chern-Simons
theories, Ph.D. Thesis [\hepth{0401185}].




\bibitem{Haro-Solodukhin-Skenderis} S. de Haro, S. N.
Solodukhin and K. Skenderis, Holographic reconstruction of
space-time and renormalization in the AdS/CFT correspondence,
\emph{Commun. Math. Phys.} \textbf{217} 595 (2001)
[\hepth{0002230}].



\bibitem{Corley} S. Corley,  A note on
holographic Ward identities, \emph{Phys. Lett.} \textbf{B484}
(2000) 141 [\hepth{0004030}].




\bibitem{Deser-Schwimmer} S. Deser and A. Schwimmer,
Geometric classification of conformal anomalies in arbitrary
dimensions, \emph{Phys. Lett.} \textbf{B309} (1993) 279
[\hepth{9302047}].




\bibitem{Imbimbo-Schwimmer-Theisen-Yankielowicz} C. Imbimbo, A. Schwimmer,
S. Theisen and S. Yankielowicz,  Diffeomorphisms and holographic
anomalies, \emph{Class. Quant. Grav.} \textbf{ 17} (2000) 1129
[\hepth{9910267}].




\bibitem{Obukhov} Yu. N. Obukhov,  Spectral geometry of the
Riemann-Cartan spacetime and the axial anomaly, \emph{ Phys.
Lett.} \textbf{B108 }(1982) 308; Spectral geometry of the
Riemann-Cartan spacetime, \emph{Nucl. Phys.} \textbf{B212} (1983)
237.




\bibitem{Urrutia-Vergara} L. F. Urrutia and J. D. Vergara,
Consistent copuling of the gravitino field to a gravitational
background with torsion, \emph{Phys. Rev. }\textbf{D44} (1991)
3882.




\bibitem{Chandia-Zanelli97} O. Chand\'{\i}a and J. Zanelli,
 Topological invariants, instatons and chiral
anomaly on spaces with torsion , \emph{Phys.Rev.} \textbf{D55}
(1997) 7580 [\hepth{9702025}].




\bibitem{Kreimer-Mielke} D. Kreimer and E. W. Mielke, Comment on
 Topological invariants, instantons and the
chiral anomaly on spaces with torsion, \emph{Phys. Rev.}
\textbf{D63}: 048501 (2001) [\grqc{9904071}]; O. Chand\'{\i}a and
J. Zanelli,  Reply to the comment by D. Kreimer and E. Mielke,
\emph{Phys. Rev.} \textbf{D63}: 048502 (2001) [\hepth{9906165}].




\end{thebibliography}
\end{document}